\documentstyle[prl,aps,psfig]{revtex}
\begin {document}

\draft

\twocolumn
[\hsize\textwidth\columnwidth\hsize\csname @twocolumnfalse\endcsname
\title{Magnetic Studies of End-Chain Spin Effects in the Haldane Gap Material
Ni(C$_{3}$H$_{10}$N$_{2}$)$_{2}$N$_{3}$(ClO$_{4}$)}

\author{G.\ E.\ Granroth\thanks{Present address:  Oak Ridge National
Laboratory,
P.O. Box 2008, MS-6393, Oak Ridge, TN 37831-6393.}, S.\ Maegawa\thanks{Present and permanent address:
Graduate School
of Human and Environmental Studies, Kyoto University, Kyoto 606-01, Japan.},
 and M.\ W.\ Meisel}
\address{Department of Physics, University of Florida, \mbox{P.O. Box 118440}, Gainesville,
 FL 32611-8440.}
 \author{J.\ Krzystek and L.-C.\ Brunel}
\address{Center for Interdisciplinary Magnetic Resonance, 
National High Magnetic Field Laboratory, 1800 East Paul Dirac Drive,
Tallahassee, FL  32306-4005.}
\author{N.\ S.\ Bell and J.\ H.\ Adair}
\address{Department of Materials Science and Engineering, University of Florida,
 \mbox{P.O. Box 116400}, Gainesville, FL 32611-6400.}
 \author{B.\ H.\ Ward, G.\ E.\ Fanucci, L.-K.\ Chou\thanks{Present address: Winbond Electronics Corp,
 No.\ 4 Creatioin Rd.\ III, Science-based Industrial Park, Hsinchu, Taiwan, R.O.C.},
 and D.\ R.\ Talham}
\address{Department of Chemistry, University of Florida, \mbox{P.O. Box 117200}, Gainesville,
 FL, 32611-7200.}

\date{\today}

\maketitle

\begin {abstract}

Electron spin resonance (ESR), at 9, 94, and 190 GHz, and magnetization
studies on polycrystalline, powder, and ultrafine powder samples of 
Ni(C$_{3}$H$_{10}$N$_{2}$)$_{2}$N$_{3}$(ClO$_{4}$) (NINAZ) have revealed several effects arising from the 
Haldane phase.  Using
the $g$ value of the end-chain spin $S$ as determined by ESR, our
results confirm that the end-chain spins are $S=\frac{1}{2}$ and show no evidence for $S=1$ end-chains.
In addition, the ESR signals reveal spectral weight consistent with a model describing interactions
between the end-chain spins on the shortest chains and between the magnetic excitations on the chains
and the end-chain spins.

\end {abstract}

\pacs{76.30.-v, 75.10.Jm, 75.50 -y}

]
Since Haldane's suggestion \cite {Haldane1} of an energy gap in the
excitation spectrum for integer spin Heisenberg antiferromagnetic chains,
significant progress has been made in experimentally and theoretically
understanding these systems.  The Haldane gap has been observed in a myriad of $S=1$ materials
\cite{Renard1,Renard2,Katsumata1,Ma1,Takeuchi1,Regnault1,Granroth3,Zheludev1},
and one $S=2$ system \cite{Granroth1}.  Theoretically, the valence bond solid
model \cite{Affleck1} provides a physically intuitive picture of  $S = 1$
antiferromagnetic chains, and one prediction is that the chains should terminate with
end-chain spins of $S=\frac{1}{2}$. The presence of  these end-chains was
identified initially by electron spin resonance (ESR) in samples of
Ni(C$_{2}$H$_{8}$N$_{2}$)$_{2}$NO$_{2}$(ClO$_{4}$) (NENP) doped with both magnetic \cite{Hagiwara1}
and non-magnetic impurities \cite {Glarum1}.  However, this
identification was challenged by Ramirez $et$ $al.$ \cite{Ramirez1} who
performed specific heat measurements on pure and doped samples of
Y$_{2}$BaNiO$_{5}$.  These results were interpreted to suggest that,
for chains up to $\sim 250$ spins long, the end-chains were $S = 1$ entities
existing in either a singlet or triplet ground state, depending on
whether the chain consisted of an even or odd number of spins.  An even/odd chain length effect is not expected for
chains longer than $\sim 50$ spins \cite{White1,Yamamoto1}, and the data of Ramirez
$et$ $al.$ \cite{Ramirez1}
have been described by $S =\frac{1}{2}$ end-chains experiencing a strong anisotropy
\cite{Desmaris1,Hallberg1}.  Although subsequent experiments \cite{Kikuchi1,Deguchi1,Ajiro1} have
been interpreted in terms of $S=\frac{1}{2}$ end-chains,
these studies have not conclusively eliminated the $S=1$ possibility.
The Haldane gap
material Ni(C$_{3}$H$_{10}$N$_{2}$)$_{2}$N$_{3}$(ClO$_{4}$) (NINAZ) \cite{Renard2} is
well-suited for the study of end-chain spins.  Our magnetization and ESR results are completely
consistent with only the presence of $S =\frac{1}{2}$ end-chains.
Furthermore, Mitra, Halperin, and Affleck (MHA) \cite{Mitra1} have described the
temperature dependence of the ESR intensity originating from the
$S = \frac{1}{2}$ end-chains, and they have made predictions about
the contributions to the linewidth arising from interactions
between the end-chain spins on the shortest chains and between the magnetic excitations on the
chains and the end-chain spins.  Although the temperature dependence of the central
peak of the ESR signal in $doped$ samples of NENP \cite{Hagiwara1,Mitra1} and TMNIN
\cite{Deguchi1} has been described by
the MHA picture, experimental evidence for the interactions
has not been reported.  Once again, due to its properties, $undoped$ NINAZ is ideally suited to
observe this phenomenon, and this Letter provides the first
experimental evidence of these fundamental interactions.

In many instances, non-magnetic dopants are used to increase the end-chain spin concentration
\cite{Glarum1,Ramirez1,Desmaris1,Kikuchi1,Deguchi1,Ajiro1,Katsumata2,Ito1}, but unfortunately,
doped samples have their limitations. For example in NENP doped beyond $\sim0.5 \%$, the
dopant no longer breaks chains by direct substitution \cite{Fujiwara1}.  Even in materials
with a non-magnetic isomorph ({\it e.g.} Ni(C$_{4}$H$_{6}$N$_{2}$)$_{2}$(C$_{2}$O$_{4}$) \cite{Kikuchi1}),
doping beyond a certain level is nonlinear with respect to the number of observed end-chain
spins.  Furthermore, dopants cause changes in the magnetic environment and may shift, split,
and/or broaden the ESR spectra.  NINAZ \cite{Renard2} circumvents doping difficulties
because it shatters while
passing through a structural phase transition at $\sim 255$ K \cite{Gadet1}, thereby producing
end-chain spins without doping \cite{Renard2,Gadet1,Chou2}.  In other words, the intrinsic
and uncontrolled shattering processes afford the distinct advantage of avoiding the
complications associated with using dopants.

\begin{figure}
\centerline{\psfig{figure=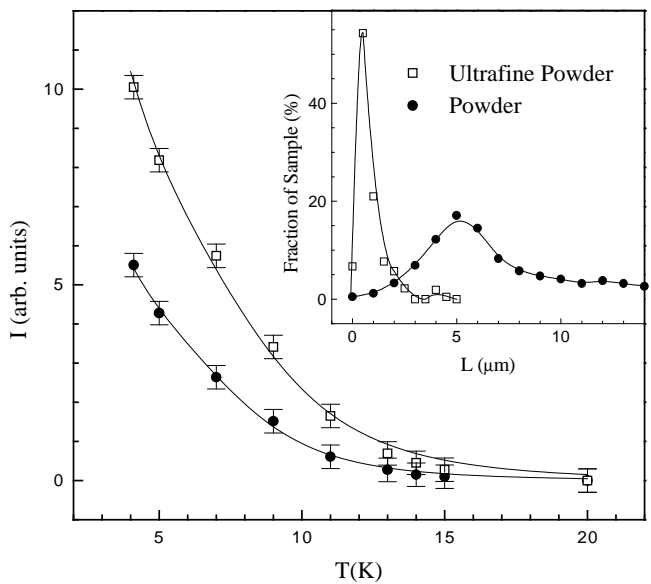,width=3.5in}}
\caption{Intensity of the 9.25 GHz ESR line for powder and ultrafine powder samples as a function of
temperature.  The solid lines are fits using Eq. (2), as described in the text. The inset shows
typical percentage distributions of particle sizes for powder and ultrafine powder
samples.  The lines are included as guides to the eye.}
\label {figure1}
\end{figure}

Two batches of NINAZ were prepared according to the
 procedure of Gadet {\it et al.}\cite{Gadet1} as modified by
 Chou \cite{Chou1}.  The room temperature crystal structure \cite{Chou1} was found to be
consistent with the one proposed by Gadet {\it et al.} \cite{Gadet1}, exhibiting a $Ni - Ni$ distance
of $a= 5.849$ \AA.  The magnetic
properties of NINAZ have been measured by several groups \cite{Renard2,Takeuchi1,Gadet1,Chou2},
and neutron scattering measurements \cite{Zheludev1} have provided a direct
measurement of the intrachain exchange energy $J = 125$ K, the Haldane gap
$\Delta = 41.9$ K, the single-ion anisotropy $D=21$ K, and the spin wave
velocity $c=2.55\times 10^{4}$ m/s.
A magnetic field, $H$, greater than 30 T is required to
close the Haldane gap\cite{Takeuchi1}, and consequently, the application of 5 T at 2 K
will not appreciably affect the Haldane state.
Since the nascent crystals shattered upon
cooling but remained oriented, these specimens are referred to as
{\it polycrystalline} samples.  To increase the number of end-chain
spins, two techniques were used to grind the samples.  Initial grindings used
a pestle and mortar to produce a sample referred to as {\it powder}.  Subsequent grindings,
using a standard ball mill, produced a sample referred to
as {\it ultrafine powder}.  Centripetal sedimentation was used to characterize the grain size
for powder and ultrafine powder samples (inset of Fig.\ \ref{figure1}).  In addition,
inductively coupled plasma mass spectrometry indicated that the level of extrinsic magnetic
impurities was no greater than a few parts per billion for any of the three sample types.
Magnetization, $M$, as a function of $H$
up to 5 T at $T=2$ K was measured using a commercial SQUID
magnetometer.
Our use of the $M(H,2 \, {\text K})$ data ensures that the Haldane state is fully developed
instead of being thermally smeared as it is in a full analysis of our
$\chi(2 \, {\text K}\leq T\leq 300 \, {\text K})$ results \cite{Granroth2}.
X-band ESR was performed with a commercial system tuned to a resonance
frequency $\nu=9.25$ GHz and operating down to $T = 4$ K.  High
frequency (93.93 and 189.87 GHz) ESR was performed at $T = 5$ K in magnetic fields up to 14 T.

\begin {figure}
\centerline{\psfig{figure=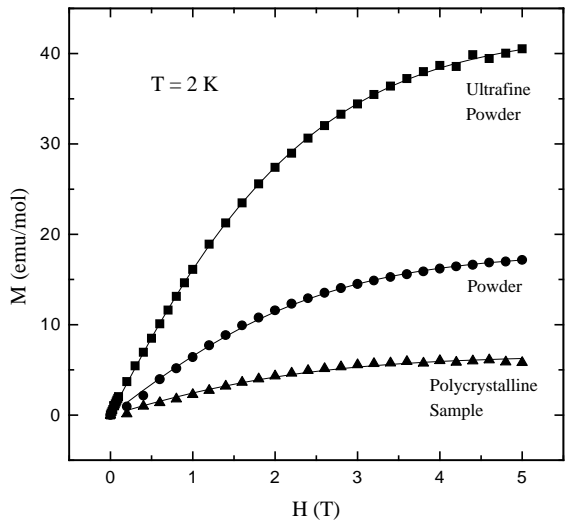,width=3.5in}}
\caption{$M(H,2 \, {\text K})$ for polycrystalline, powder, and
ultrafine powder samples. The experimental uncertainties are given by the size
of the data points. The solid lines represent fits to the sum of two
Brillouin functions, Eq. (1), as described in the text.}
\label {figure2}
\end {figure}

Figure\ \ref{figure2} shows the $M(H,2\, {\text K})$ data for polycrystalline, powder, and ultrafine powder
samples.  The solid lines are fits using
\begin{eqnarray}
M(H,T)&=&N_{A}g\mu _{B}\left[\left(1/2\right)N_{1/2}B_{1/2}
(g\mu _{B}H/k_{B}T)\right. \nonumber \\
 &&\left. + N_{1}B_{1} (g\mu _{B}H/k_{B}T)\right],
 \end{eqnarray}
where $B_{1/2}(g\mu _{B}H/k_{B}T)$ and $B_{1}(g\mu _{B}H/k_{B}T)$ are the Brillouin functions, and
$N_{1/2}$ and $N_{1}$ are the
concentrations of $S=\frac{1}{2}$ and $S=1$ spins.
For the fits, $g=2.174$ was determined from ESR measurements, leaving $N_{1/2}$ and $N_{1}$
as the only free parameters since no measurable amounts of other spin values are reasonably
expected or were observed by any ESR frequency.  The results of the fits are given in
Table \ref{table1} and demonstrate that the end-chain spins are predominately $S=\frac{1}{2}$ and
that only trace amounts of $S=1$, consistent with the presence of some uncoupled Ni$^{2+}$, exist in
any of the samples.  If there were contributions from singlet and triplet ground states, 
one would expect significantly more $S=1$ than
$S=\frac{1}{2}$ spins.  Furthermore, our high frequency ESR studies did not reveal any $S=1$
resonances.  Consequently, we conclude all the end-chain spins are $S=\frac{1}{2}$.

For our 9.25 GHz ESR work, the measured derivative spectra were integrated at a variety of temperatures.
In the MHA description \cite{Mitra1},
the temperature dependence of the intensity of the central peak is given by the fraction of
finite-length chains in their ground state and may be written as
\begin{eqnarray}
I(T)&=&I(0)\tanh \left( \frac{h \nu}{2k_{B}T}\right) \nonumber \\
& &\times \frac{\exp \left[-\left( \pi^{-1/2}L_{min}\lambda_{T}^{-1} -1\right)
e^{\frac{-\Delta}{k_{B}T}}\right]}{1+\pi^{-1/2}L_{0}\lambda_{T}^{-1}e^{\frac{-\Delta}{k_{B}T}}},
\label{mitrafit1}
\end{eqnarray}
where $\nu = 9.25$ GHz, $\lambda_{T}= \hbar c (2k_{B}T \Delta)^{-1/2}$,
$L_{0}$ is the average chain length, and
$L_{min}$ is the minimum chain length.
Since the Haldane gap $\Delta \approx 42$ K is independently known,
$L_{0}$ and $L_{min}$ are the only parameters, and
$L_{min}$ is constrained by the strength of the interaction between the $S=\frac{1}{2}$ spins on
the shortest chains \cite{Mitra1}.  Figure\ \ref{figure1} shows the temperature dependence of
the intensity for the powder and
ultrafine powder specimens and the results of fits using Eq.\ \ref{mitrafit1} with
$L_{min}=60\pm20$ sites.  Excellent agreement exists between the data and the fits
when $L_{0} = 1590$ $\pm$ $50$ sites ($\sim$ $0.9$ $\mu $m) for the
powder and $910$ $\pm$ $50$ sites ($\sim 0.5$ $\mu m$) for the ultrafine powder.
Comparison with the particle size analysis (inset of Fig.\ \ref{figure1}) confirms that the
pestle and mortar grinding process only slightly reduces the average chain length when compared
to the polycrystalline material \cite{Chou2}, whereas ball milling produces chains of a length
consistent with the particle size.  Finally, the values of $L_{0}$ may be compared to the ones
obtained from the magnetization results.  Assuming a Poisson distribution, the results listed in
Table I suggest $L_{0} \approx 2300$ sites for the powder and $\approx 1100$ sites for the
ultrafine powder.  Within reasonable expectations for the model and analysis, these values are consistent
with the aforementioned ones.

\begin{figure}
\centerline{\psfig{figure=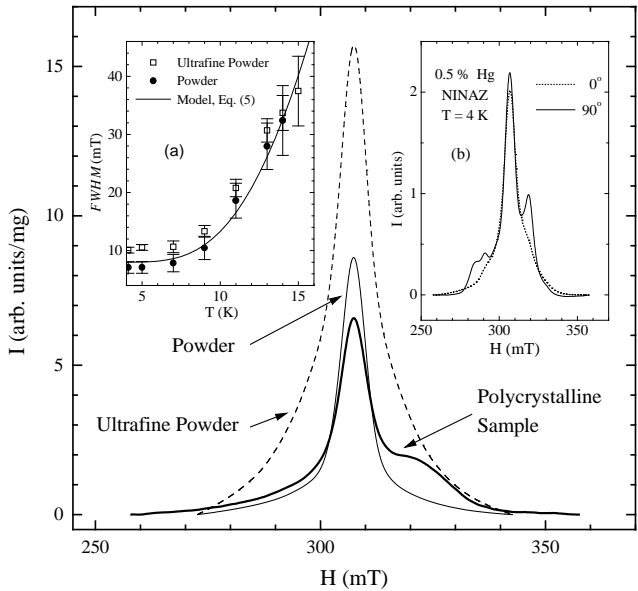,width=3.5in}}
\caption {Typical 9.25 GHz ESR line shapes for polycrystalline, powder, and ultrafine
powder samples.  Inset (a) shows the temperature dependence of the $FWHM$.  The solid line
represents a fit using Eq. (5), as described in the text.
Inset (b) shows the orientational
dependence of a $0.5\%$ Hg doped sample, and this behavior
is absent in the undoped polycrystalline specimen.}

\label{figure4}

\end{figure}

In Fig.\ \ref{figure4}, typical 9.25 GHz ESR lines at $T=4$ K are plotted, showing that the full width
at half maxima ($FWHM$) for the polycrystalline, powder, and
ultrafine powder samples are $\approx 8$ mT, $\approx 7$ mT, and $\approx 10$ mT,
respectively.  The temperature dependencies of the linewidths are shown in Fig.\ 3a.
For $T\leq 7$ K, the $FWHM$ is temperature independent, and this limit is governed by the interactions
between the $S=\frac{1}{2}$ end-chains on the shortest chains.  Following MHA \cite{Mitra1},
this interaction may be roughly estimated as
\begin{equation}
\epsilon=\Delta \exp (-L_{min}/\xi),
\end{equation}
where $\xi$ is the
correlation length.  When $\Delta \approx 42$ K, $\xi \approx 6$
\cite{White1,Yamamoto1}, and $L_{min} = 50$, then the resultant energy is
$\epsilon \approx$ 10 mK $\approx$ 8 mT [30].

When magnetic excitations are present on the chains, there are two mechanisms \cite{Mitra1} by which
these bosons may influence the linewidth.
Firstly, bosons changing energy levels (via interactions with the end-chain spins)
could affect the linewidth, where this change in
energy is quantized in units of
\begin{equation}
\delta E \approx \frac{(\hbar \pi c)^{2}}{2 \Delta L^{2}}.
\end{equation}
Using the values quoted earlier and taking $L = L_{0}$, $\delta E \approx 5$ mT for the powder and
$\approx 12$ mT for the ultrafine powder.  These values of $\delta E$ are about the size of the $FWHM$, and
consequently, these interactions contribute to the linewidth once a chain acquires a boson.  The temperature
dependence of the linewidth near the central peak has been derived by MHA [22], so the $FWHM$ may be modeled
by
\begin{equation}
FWHM = \epsilon + \Lambda \: T \exp(-\Delta/k_{B}T),
\end{equation}
where $\Lambda$ is a parameter which is beyond the scope of the present MHA model.
The temperature dependence
of the $FWHM$ is reasonably reproduced when $\epsilon = 8$ mT and $\Lambda = 35$ mT/K, as shown in Fig.\ 3a.
In other words, the
major contribution to the $FWHM$, at $T \leq 7$ K, comes from the end-chain spins on the shortest chains
interacting with each other and, at $T \geq 7$ K, arises from the interactions between the magnetic
excitations and the end-chain spins.
Finally, the other possible broadening mechanism arises from a small change in the energy of a boson that
experiences a phase shift when interacting with an end-chain spin.
This effect is significantly weaker than the exchange of energy $\delta E$ and,
consequently, is not detectable in the present measurements.

Another effect shown in Fig.\ 3 is the multiple
ESR peaks of the polycrystalline sample. The polycrystalline line is the superposition of two
strong peaks, the main one at $307.3$ $\pm$ $0.2$ mT and the other at $318.4$ $\pm$ $0.4$ mT.  While the
$307$ mT line is due to the $S = \frac{1}{2}$ end-chains,
the intensity of the $318$ mT peak depends both on the number of times
the sample is cycled through the structural transition as well as the crystal
used, suggesting that this peak is a consequence of the shattering
process \cite{Granroth2}.
Since $S=1$ spins and extrinsic impurities are not possible explanations, one may consider
changes in the magnetic environment due to chemical differences at surface sites or a
dipolar interaction with a nearby end-chain spin.
These possibilities may be examined by rotating the sample, but the lines observed with our undoped
materials were rotationally invariant.  On the other hand, in addition to the main ESR line which
did not move when the sample was rotated, 0.5\% Hg doped polycrystalline specimens exhibited lines 
which were
dependent on orientation (see Fig.\ 3b).  These results indicate different magnetic sites were within our
resolution and illustrate a reason why linewidth measurements of other systems
failed to detect the presence of interactions 
($i.e.$ doping broadens the main line and adds other extrinsic ones).

In conclusion, our studies of the end-chains of NINAZ have experimentally
confirmed fundamental theoretical predictions about quantum spin chains.
Firstly, our magnetization and ESR measurements identify the presence of $S=\frac{1}{2}$ end-chains
(see Figs. 1 - 3) and detect no evidence of $S=1$ end-chain spins.
Secondly, our ESR studies reveal spectral weight arising from the interaction between
the end-chain spins on the shortest chains and between the magnetic excitations on the chains and the
end-chain spins.

We thank W.\ W.\ Kim, P.\ J.\ C.\ Signore, M.\ Escobar, and R.\ Russell
for contributions to this work.
We have enjoyed communications or discussions with J. H. Barry, T. M. Brill, E.\ Dagotto, A. Feher, 
K. Hallberg, A. K. Hassan, J.\ K.\ Ingersent,
Th.\ Jolic\oe ur, K.\ Majumdar, S.\ E.\ Nagler, M. Orend\'{a}\u{c}, C.\ Saylor, F.\ Sharifi,
and A.\ Sikkema.  This work was
supported, in part, by funding from the NSF through DMR-9200671 (M.W.M.),
DMR-9530453
(D.R.T.), and the NHMFL.
\\

$^{*}$Present address:  Oak Ridge National Laboratory,
P.O. Box 2008, MS-6393, Oak Ridge, TN 37831-6393.

$^{\dagger}$Present and permanent address:
Graduate School
of Human and Environmental Studies, Kyoto University, Kyoto 606-01, Japan.

$^{\ddagger}$Present address: Winbond Electronics Corp,
 No.\ 4 Creatioin Rd.\ III, Science-based Industrial Park, Hsinchu, Taiwan, R.O.C.

\newpage
\begin{table}
\begin{tabular}{lccc}
&Polycrystalline Sample&Powder&Ultrafine powder\\
$N_{1/2}$& $1000\pm 60$&$2840\pm 50$&$6280\pm 50$\\
$N_{1}$&$20\pm 10$&$30\pm 9$&$140\pm 8$
\end{tabular}

\caption{$N_{1/2}$ and $N_{1}$ (in ppm) of $S=\frac{1}{2}$ and
$S=1$ spins in the polycrystalline, powder, and
ultrafine powder samples as deduced by fitting the $M(H,2\, {\text K})$ data (Fig.\ \ref{figure2}) to Eq.\ (1).}
\label{table1}
\end{table}

\end{document}